\documentclass[prd,nofootinbib]{revtex4-1}

\usepackage{amssymb}
\usepackage{amsfonts}
\usepackage{amsmath,color}
\usepackage{bm}

\usepackage{hyperref}

\hypersetup{
    colorlinks=true,
    linkcolor=red,
    citecolor=red,
    filecolor=magenta,      
    urlcolor=cyan,
    pdftitle={Galactic Dynamics in General Relativity: the Role of Gravitomagnetism},
    pdfpagemode=FullScreen,}

\def\mb#1{\mathbf{#1}}

\def\ber{\begin{eqnarray}}
\def\eer{\end{eqnarray}}
\def\beq{\begin{equation}}
\def\eeq{\end{equation}}

\def\rmd{{\rm d}}

\def\ed{\end{document}}

\newcommand{\ppar}[2]{\frac{\partial #1}{\partial #2}}
\newcommand{\pparo}[2]{\frac{\partial^{2} #1}{\partial #2}}

\begin{document}

\author{Matteo Luca Ruggiero}
\email{matteoluca.ruggiero@unito.it}
\affiliation{Dipartimento di Matematica ``G.Peano'', Universit\`a degli studi di Torino, Via Carlo Alberto 10, 10123 Torino, Italy}
\affiliation{INFN - LNL , Viale dell'Universit\`a 2, 35020 Legnaro (PD), Italy}
\author{Antonello Ortolan}
\email{antonello.ortolan@lnl.infn.it}
\affiliation{INFN - LNL , Viale dell'Universit\`a 2, 35020 Legnaro (PD), Italy}
\author{Clive C. Speake }
\email{c.c.speake@bham.ac.uk}
\affiliation{School of Physics and Astronomy, University of Birmingham, Edgbaston, Birmingham B15 2TT, UK}

\date{\today}

\title{Galactic Dynamics in General Relativity: the Role of Gravitomagnetism}

\begin{abstract}
It is a well known fact that, in the absence of Dark Matter, the observation of the rotation curves of galaxies cannot be explained in terms of Newtonian gravity. Rotation curves become flat in the outer regions, in contrast to what is expected according to Keplerian motion. {Far from the galactic center, the gravitational field is supposed to be weak enough so  we expect to be able to use Newtonian Gravity;  however, even in the weak-field approximation, there are general relativistic effects without a Newtonian counterpart, such as the gravitomagnetic effects originating from mass currents.} Using the   gravitoelectromagnetic approach to the solution of Einstein equations in the weak-field and slow-motion approximation, we discuss some simple arguments that suggest the surprising result that gravitomagnetic effects may have a relevant role \textcolor{black}{in better understanding the impact of Dark Matter on galactic dynamics.} In addition, treating matter as a  fluid of dust, we  study the influence of  post-Newtonian effects on the fluid vorticity.
\end{abstract}

\maketitle

\section{Introduction}\label{sec:intro}

Thanks to the pioneering work of \citet{rubin1978extended} and taking into account subsequent analyses \cite{sofue2001rotation}, we know that the rotation curves of galaxies become  flat far from the center: the observed non-Keplerian features contradict what is expected on the basis of a Newtonian gravity analysis. The observation of flatness of rotation curves is one element in the case supporting the existence of \textit{dark matter}   
\cite{strigari2013galactic,amendola2018cosmology}. It is also worth pointing out that an \textit{ad hoc} modification of Newtonian dynamics, the so-called MOND has also been proposed by \citet{Milgrom:1983ca} to explain the flatness problem.\\
\indent At first sight, more than 100 years after the publication of General Relativity (GR), it could appear surprising that Newtonian rather than Einsteinian physics is applied to study the gravitational dynamics of galaxies. The motivations are apparently sound, since the stars' velocities are small compared to the speed of light and, moreover, in the outer regions of the galaxies far from the center, the gravitational field is weak. However, a subtler analysis suggests a different perspective. In fact, while we know that Newtonian gravity is a good approximation to GR, we should remember that there are general relativistic effects without Newtonian counterparts: for instance, gravitational waves and the effects produced by mass currents, the so-called \textit{gravitomagnetic field} which gives rise to the Lense-Thirring effect \cite{iorio2011,pfister2014gravitomagnetism}. Recently, gravitomagnetic effects due to passage of a plane gravitational wave have been investigated \cite{Ruggiero_2020,Ruggiero_2020b,Ruggiero:2021qnu}. In both  cases, GR introduces new phenomena that go beyond  corrections to known Newtonian effects.\\ 
\indent Based on these motivations, there have been some attempts to use suitable solutions of GR field equations to model galaxies and, in this framework, the flatness problem is analysed without requiring extra unknown sources.  We refer, for instance, to the works published by \citet{Cooperstock:2006dt} and \citet{Balasin:2006cg}. In both cases, a solution is obtained with the assumptions that (i) the system is axially symmetric and stationary, (ii) matter can be treated as dust {(pressureless and perfect - hence inviscid - fluid  interacting only gravitationally)}.  The model proposed by \citet{Cooperstock:2006dt} has been questioned  (see e.g. \citet{Cross:2006rx,Menzies:2007dm}), while the model developed by \citet{Balasin:2006cg} even though it develops singularities, has been recently reconsidered by \citet{crosta2020testing} and generalized by \citet{Astesiano:2021ren}. \textcolor{black}{The corrections due to the galactic rotation on the velocity of circular orbits were calculated by \cite{Vogt:2005hn}, using an approximate form of the Kerr solution}. The role of post-Newtonian corrections in galactic dynamics was emphasized by \citet{Ramos-Caro:2012ren} while, in a recently published paper, \citet{ludwig2021galactic}
 developed a model of galactic dynamics based on the equations that govern the motion of a weakly relativistic perfect fluid. 
All these research efforts share the idea that the gravitomagnetic effects, originating from mass currents, have a relevant impact on the observed galactic dynamics. Consequently, if these purely non-Newtonian effects are properly taken into account, there is perhaps no need to invoke dark matter to explain observations or {it is likely that} the impact of dark matter should be reconsidered.

The purpose of this paper is to discuss, from a quite general perspective, the role of gravitomagnetic effects in the study of galactic dynamics. {This might help to shed new light on Dark Matter, which plays a central role not only in this context, but in many domains of modern cosmology.} To this end, we will not resort to specific GR solutions or solve the fluid equation coupled to the gravitational field using specific mass density models. We will just assume that, in the outer regions of the galaxies, a weak-field and slow-motion approximation can be used to describe the gravitational field acting upon stars, which can themselves be treated as test masses and whose motion can be studied using the geodesic equation: {in particular, we will consider the stars as  dust.}  In addition, we will analyse  the limits of the Maxwellian analogy, since new terms that were neglected in previous analysis have an effective impact on the fluid vorticity and can be relevant for astrophysical systems.\\

\section{Gravitoelectromagnetism in a nutshell} \label{sec:GEM}

The solution of Einstein field equations in weak-field and slow-motion approximation,  leads to the gravitoelectromagnetic analogy; in particular, it is possible to write a set of Maxwell-like equations,  where the mass density and current play the role of the charge density and current, respectively \cite{Mashhoon:2001ir,ciufolini1995gravitation,Costa:2012cw}. Accordingly, if we neglect in the metric tensor terms that are $O(c^{-4})$, the line element can be written\footnote{Bold symbols like $\mb v$ refer to space vectors, in particular $\mb x=(x,y,z)$ is the position vector at location P(x,y,z); the spacetime signature is $(-1,1,1,1)$;  Latin indices run  from 1 to 3; {$G$ is the gravitational constant and $c$ is the speed of light in vacuum.} } as \cite{Ruggiero:2002hz,Mashhoon:2003ax}
\begin{equation}
\mathrm{d} s^2= -c^2 \left(1-2\frac{\Phi}{c^2}\right)\rmd t^2 -\frac4c ({\mathbf A}\cdot \rmd {\mathbf x})\rmd t
  + \left(1+2\frac{\Phi}{c^2}\right)\delta_{ij}\rmd x^i \rmd x^j\, \label{eq:weakfieldmetric1}
\end{equation}
where, in stationary conditions, the gravitoelectric  ($\Phi $) and gravitomagnetic  ($\mathbf A$) potentials are solutions of Poisson equations 
\begin{eqnarray}
\nabla^{2} \Phi&=&-4\pi G\, \rho, \label{eq:poisson01} \\
\nabla^{2} \mb A & =& -8\pi G\, \mb j,  \label{eq:poisson02}
\end{eqnarray}
in terms of the mass density $\rho_{  }$ and current $\mathbf j$ of the sources. The analogy with electromagnetism is strengthened  by the geodesic equation. In fact, {in stationary conditions,} if we define the gravitoelectric ($\mb E$) and gravitomagnetic ($\mb B$) fields 
\beq
\mb E=-\bm \nabla \Phi, \quad  \mb B= \bm \nabla \wedge \mb A, \label{eq:solgemEB1}
\eeq
the spatial components of the geodesic equation, up to linear order in $\frac{|\mb v|}{c}$,   can be expressed in analogy with the Lorentz-like force equation acting upon a test mass $m$
\beq
m\frac{\rmd {\mathbf v}}{\rmd t}=-m{\mathbf E}-2m \frac{{\mathbf v}}{c}\wedge {\mathbf B}. \label{eq:lor2}
\eeq
{Notice that the hypothesis of stationarity is used not only in the definition of the gravitoelectromagnetic fields (\ref{eq:solgemEB1}), but also in the expression of the geodesic equation; if we relax the stationary conditions and define the gravitoelectromagnetic fields}
\beq
\mb B= \bm \nabla \wedge \mb A, \quad \mb E= -\bm \nabla \Phi-\frac{2}{c} \ppar{\mb A}{t}, \label{eq:defEtime}
\eeq
the above equation (\ref{eq:lor2}) becomes \cite{bini2008gravitational,Costa:2012cw,Ruggiero:2021uag} 
\beq
m\frac{\rmd {\mathbf v}}{\rmd t}=-m{\mathbf E}-2m \frac{{\mathbf v}}{c}\wedge {\mathbf B}-3m\frac{\mb v}{c} \frac{\partial \Phi }{c\partial t} \label{eq:lor222}
\eeq
In particular, we notice that the  last term in Eq. (\ref{eq:lor222})  is non-Maxwellian and, hence, breaks the gravitoelectromagnetic analogy: we will discuss its impact in what follows.\\
The above approach is aimed at emphasizing the similarity between electromagnetism and general relativity in the weak-field and slow motion approximation. In analogy with the electric field of a point charge,  $\Phi$ differs by a minus sign from the actual Newtonian potential of point mass $M$, $\Psi=-\frac{GM}{|\mb x|}$. In what follows, it will be useful to make a comparison with Newtonian quantities: so, to this end, we introduce the gravitational field $\mb g=-\bm \nabla \Psi=-\mb E$ and the  Newtonian potential $\Psi=-\Phi$.\\
On the other hand we compute the gravitomagnetic field in the same way as in MaxwellÕs equations. However the contribution to the Lorentz force from the gravitomagnetic field is multiplied by a factor (-2) (see e.g. Eq. (\ref{eq:lor2})).


\section{The Generalized Vorticity} \label{sec:genvor}

Using the gravitoelectromagnetic analogy, we can define some properties of a weakly relativistic dust. To this end,  we remember that when we have fluids and electromagnetic fields it is possible to introduce the \textit{generalized vorticity}  $\bm \Omega_{E}=\bm \nabla \wedge \bm{\mathcal P}$ starting from the canonical momentum $\bm{\mathcal P}=m\mb v+\frac q c \bm{\mathcal A}$, {where $q$ is the electromagnetic charge and $\bm{\mathcal A}$ the electromagnetic vector potential.} Similarly,  for a weakly relativistic dust we may define $\mb P=\mb v-\frac 2 c \mb A $, that  is the canonical momentum of the fluid element per unit mass \cite{Mashhoon:2003ax}; for the sake of simplicity, here we focus only on gravitational effects and do not consider electromagnetic ones. Using these definitions,  we may write the momentum equation (\ref{eq:lor2}) in the form:
\beq
\ppar{\mb P}{t}+\left(\bm \nabla \wedge \mb P \right) \wedge \mb v=\bm \nabla\left(\Phi-\frac 1 2  v^{2} \right).  \label{eq:mom123}
\eeq
In addition, we may define the \textit{gravitational generalized vorticity} $\bm \Omega_{G}=\bm \nabla \wedge \bm P=\bm \Omega+2\bm\Omega_{L}$ in terms of the fluid vorticity $\bm \Omega= \bm \nabla \wedge \mb v$, and the  gravitomagnetic Larmor frequency $\bm\Omega_{L}=-\frac 1 c \mb B$ \cite{Mashhoonspin}. Due to the formal analogy with the electromagnetic case (see e.g. \citet{mahajan2010twisting}),  the rate of change of circulation
\beq
\Gamma=\oint_{L} \mb{ P} \cdot d\mb l = \int_{S} \bm \Omega_{G} \cdot d\mb S=\int_{S} \bm \Omega\cdot d\mb S+\int_{S} 2\bm \Omega_{L}\cdot d\mb S=\Gamma_{N}+\Gamma_{GM} \ \label{eq:can2}
\eeq
along a closed loop $L$, which is the contour of the surface $S$, is zero.  Hence, we see that a gravitomagnetic field is indistinguishable from a vorticity field, and thus it contributes to establish a rotational fluid. The generalized  gravitational vorticity is conserved, and cannot emerge from a zero initial value; accordingly, the Newtonian circulation $\Gamma_{N}$  is not conserved  due to the presence of the gravitomagnetic term $\Gamma_{GM}$: the latter, is simply related to the gravitomagnetic flux and its presence can be relevant in coalescence of compact binaries or accretion phenomena around black holes \cite{bhattacharjee2021gravitomagnetic,shapiro1996gravitomagnetic}.  The presence of the gravitomagnetic field can explain rotational motion in a inviscid fluid as a consequence of  the definition of generalized vorticity.  Moreover, the generalized gravitational helicity
\beq
H=\int_{V} \mathbf{P} \cdot \bm \Omega_{G} dV  \label{eq:can3}
\eeq
is conserved (see e.g. \citet{mahajan2003temperature,Alves:2017zjt} and references therein; see also \citet{Bini:2021gdb} for the gravitomagnetic helicity).  Again, the Newtonian helicity is not conserved, which means that the gravitomagnetic field modifies the topology of the line of force of the Newtonian fluid.\\
\indent If we want to analyse our fluid of  stars using this approach, we must remember the  gravitomagnetic field experienced by a test mass at a given location depends both on matter moving \textit{inside} and \textit{outside} its orbit \cite{ciufolini2003gravitomagnetic,ruggiero2016gravitomagnetic}. Consequently,   it is necessary to treat the system as a whole, and to look for a self consistent solution of the gravitational and fluid equations.
To this end, in stationary conditions the momentum  equation (\ref{eq:mom123}) can be written in the form
\textcolor{black}{
\beq
\bm \nabla \left(\frac{v^{2}}{2} + \Psi \right)=\mb v \wedge \bm \Omega_{G}. \label{eq:mom2a}
\eeq}
This equation, together with the source equations (\ref{eq:poisson01})-(\ref{eq:poisson02}) can be used to describe the equilibrium of the dust: so, starting from the knowledge of the mass density profile, it is possible to use the above  equations, to obtain $\mb v, \mb A, \Psi$ {(remember that the mass current is defined by $\mb j=\rho \mb v$, in terms of the mass density and fluid velocity). }

 Actually, this is what is done in the paper by \citet{ludwig2021galactic}.

\subsection{Non-stationary conditions} \label{sec:nonstationary}

The gravitoelectromagnetic analogy has however some limitations, which are important to emphasize, in particular if we consider the non-stationary state, that is in the evolution of the fluid of dust. In fact, we noticed that in these conditions the momentum equation (\ref{eq:lor222}) has a non-Maxwellian term which is missing in previous analysis (see e.g. \citet{ludwig2021galactic}) but it is reasonable that it cannot be neglected in an evolving scenario. In fact, at a distance $r$ from a point-like source of mass $M$, moving with speed $v_{s}$, its magnitude is $\displaystyle \left|  \frac{v}{c} \frac{\partial \left(\Phi \right)}{c\partial t} \right| \simeq \frac{Mv v_{s}}{c^{2}r^{2}}$:  accordingly,  it is of the same order as the translational gravitomagnetic field of the source. The non-Maxwellian term can be neglected if we assume that the source is at rest,  but in evolving  scenarios  it cannot be neglected. The presence of the non Maxwellian term, then, leads to the \textit{non conservation} of the generalized gravitational vorticity and the corresponding helicity. Moreover, starting from Eq. (\ref{eq:lor222}) and using the continuity equation, it is possible to show that  $\bm \Omega_{G}$ satisfies the following vorticity equation
\beq
\frac{\rmd { }}{\rmd t}\left(\frac{\bm \Omega_{G}}{\rho} \right)=\left(\frac{\bm \Omega_{G}}{\rho} \cdot \bm \nabla \right) \mb v+\frac{1}{\rho}\bm{\mathcal S} \label{eq:Omegagen1}
\eeq
where   $\displaystyle \bm{\mathcal S}=-\frac{3}{c^{2}}\bm \nabla\left(\frac{\partial \Phi}{\partial t} \right) \wedge \mb v-\frac{3}{c^{2}}\frac{\partial \Phi}{\partial t} \ \bm \nabla \wedge \mb v$.  
This is a new result that generalizes what was obtained in previous works \cite{ludwig2021galactic}, and emphasises the importance of all post-Newtonian effects (and not only of the gravitomagnetic ones) in determining the evolution of a fluid element. Indeed, the presence of the gravitomagnetic effect \textit{and} of the non-Maxwellian term may produce vorticity in an initially irrotational motion. It is expected that this may have implications in galaxies dynamics and, more in general, in the evolution of astrophysical systems. \\

\section{The Newtonian case} \label{sec:newton}

 Initially we examine the possibility of describing an isolated, axially symmetric and rotating dust solution using only Newtonian physics. The limitations of this approach in comparison with GR were pointed out by \citet{bonnor1977rotating} in 1977 and we briefly discuss his argument. We consider dust particles  steadily rotating around a symmetry axis: we use a reference frame, with origin $O$,  and choose cylindrical coordinates $\{r,\varphi,z\}$ within it, such that $z$ is the rotation axis; $\mb u_{r}, \mb u_{\varphi}, \mb u_{z}$ are the unit vectors. If $\bm \omega=\omega \mb u_{z}$ is the rotation rate and $\mb x$ is the position vector of a dust particle, we may write its velocity in the form
\beq
\mb v=\bm \omega \wedge \mb x, \label{eq:vel1}
\eeq 
where we suppose that $\bm \omega$ can be a function of ${r}$ and ${z}$.
The momentum equation for a dust cloud acted upon by the gravitational field is: 
\beq
\ppar{\mb v}{t}+\left(\mb v \cdot \bm \nabla \right) \mb v=-\bm \nabla \Psi, \label{eq:momN1}
\eeq
where $\mb v$ is the velocity and again $\Psi$ the Newtonian potential, and we have used the convective derivative $\displaystyle \frac{\rmd { }}{\rmd t}=\ppar{}{t}+\mb v \cdot \bm \nabla$. The continuity equation may be  written as
\beq
\ppar{\rho}{t}+\bm \nabla \cdot (\rho \mb v)=0 .\label{eq:contN1}
\eeq
In stationary conditions ($\ppar{}{t}=0$), the above equations become
\begin{eqnarray}
\left(\mb v \cdot \bm \nabla \right) \mb v=-\bm \nabla \Psi, \label{eq:mom11} \\
\bm \nabla \cdot (\rho \mb v)=0, \label{eq:cont1}
\end{eqnarray}
The gravitational potential and the mass density are related by the Poisson equation
\beq
\nabla^{2} \Psi=4\pi G \rho. \label{eq:poisson1}
\eeq
It is useful for what follows to write the following mathematical identities 
\beq
\displaystyle \nabla^{2} \Psi=\frac 1 r \ppar{}{r}\left({r}\ppar{\Psi}{r} \right)+\pparo{\Psi}{z^{2}},\label{eq:Laplacian1}
\eeq
and
\beq
\left(\mb v \cdot \bm \nabla \right) \mb v=\left(\nabla \wedge \mb v\right)\wedge\mb v+\frac 1 2 \nabla \left( \mb v \cdot \mb v \right). \label{eq:identity1}
\eeq
Since, according to Eq. (\ref{eq:vel1}) $v_{z}=0$ everywhere,  from the momentum equation (\ref{eq:momN1}) we obtain 
\beq
\ppar{\Psi}{z}=0. \label{eq:Phiz}
\eeq
In other words there is no gravitational force along the rotation axis acting on the dust particles. As a consequence, by differentiating the Poisson equation (\ref{eq:poisson1}), we eventually deduce that  there is no density gradient parallel to the rotation axis:
\beq
\ppar{\rho}{z}=0, \label{eq:rhoz}
\eeq
Eventually, on taking into account axial symmetry and using Eq. (\ref{eq:identity1}), Eq. (\ref{eq:momN1}) turns out to be
\beq
\omega^{2}r=\ppar{\Psi}{r}, \label{eq:Phir}
\eeq
which means that also $\omega$ is independent of $z$, and the motion of the dust particles is the same in every plane $z=$constant. In summary: \textit{in Newtonian gravity, stationary, axially symmetric motion of dust is necessarily cylindrically symmetric} and, moreover,  the motion is the same in every plane orthogonal to the symmetry axis. Hence, no compact or finite dust object can exist in Newtonian gravity in the given symmetry conditions. As Bonnor pointed out, things are quite different in general relativity where ``a non-Newtonian force, arising from the spin of the central body, permits non-equatorial circular  orbits''. Actually, as we are going to see below, this can be explained in terms of a gravitomagnetic force acting on  moving masses.\\ 
\indent In several papers and textbooks pertaining to galactic dynamics,  it is maintained that for those that are rotationally supported, such as spirals or irregular, relation (\ref{eq:Phir})  provides a link between rotation curves and gravitational potential \cite{mcgaugh2016radial}.  The analysis of the rotation curves is based on the following hypothesis: since the luminous mass density is rapidly decreasing from the center, the galaxy is modelled as a point mass, and it is then expected that stars far away from the center move as test masses around a point-like mass, just like the planets around the Sun. Accordingly, Newtonian dynamics suggests that velocities should decrease, which is not observed, since rotation curves remain flat. Hence, it is expected that \textit{extra, non-visible matter} is present and acts upon stars, thus increasing their velocity,   the so-called dark matter.  This is different from the Newtonian analysis of a self-consistent dust solution that we have considered,  since in the standard approach it is supposed that there is a compact mass distribution, which constitutes the great part of the galaxy mass content, and distant stars move like test particles around it.\\
\indent {Our discussion of Newtonian solutions has been limited to models with smoothly varying functions of density. However we note that  the solutions proposed by  \citet{Kuzmin} and   \citet{Toomre} are based on infinitely thin discs of finite radial extent but therefore feature unphysical discontinuities in the axial gravity field. Other models having mass distributions that have a finite axial extent  have been found by \citet{Miyamoto}, but these cannot form time-independent structures as suggested by the above analysis. {{The stability of these structures is generally attributed to the dispersive random component of velocity 
and the deepening of the potential well of the galaxy itself by the Dark Matter Halo\cite{ostriker}}. } Axial forces are required to maintain equilibrium against the axial gravity field as we will see below.}\\
Starting from the Poisson equation (\ref{eq:poisson1}), and substituting in it Eq. (\ref{eq:Phir}), we obtain 
\beq
2\omega^{2}+2\omega\ppar{\omega}{r}r=4\pi G \rho. \label{eq:rhoomega}
\eeq
This equation locally relates the matter density to the rotation rate and its derivative: in the regime $v=\omega r \simeq \mathrm{constant}$, that is in the flat region of the rotation curves, we have $\ppar{\omega}{r}r+\omega \simeq 0$, hence we get $\rho=0$ from (\ref{eq:rhoomega}). As a consequence, in the framework of Newtonian gravity the flat velocity profile for the infinite dust cylinder is not allowed \cite{cooperstock2016weak,cooperstock2016power}. Stated in a different way, using a purely Newtonian approach, it is not clear how to relate the matter density with the rotation rate in the flat zone. The situation is quite different if we resort to General Relativity. \\

\section{The Einsteinian case} \label{sec:einstein}

Let us now  extend the  analysis of an axially symmetric and rotating dust solution using the general relativistic approach; in particular, {dust}   will be coupled to the  gravitational field described in terms of the gravitoelectromagnetic analogy.  Before entering into details, we describe the underlying hypotheses. A galaxy as a whole is a complex object in which very different gravitational fields are present: think of the Milky Way, with a supermassive black hole \cite{Ghez:1998ph} at its center, and the disk which extends up to 50 kpc. Then, it is reasonable to say  that the dust approximation cannot be used to describe the galactic center, but in the outer regions it is expected that stars can be thought of as test particles, moving in a background spacetime, that can be considered a solution of Einstein equations in weak-field and slow-motion approximation, with the additional requirements of axial symmetry and stationarity. Accordingly, this solution can be expressed in the form (\ref{eq:weakfieldmetric1}).  We point out that for our purposes  it is not necessary to know the explicit form of the spacetime metric: we just want to focus on the role of the gravitomagnetic effects. 
Stars move along geodesics defined by the Lorentz-like equation 
\beq
\mb a=\mb g-2 \frac{{\mathbf v}}{c}\wedge {\mathbf B}, \label{eq:lor221}
\eeq
where their acceleration can be written as a convective derivative of velocity  $\mb a= \frac{\rmd {\mathbf v}}{\rmd t}=\ppar{{\mathbf v}}{t}+\mb v \cdot \bm \nabla{\mathbf v}$. 
This is the momentum equation for the dust; the difference  with respect to the corresponding  Newtonian equation (\ref{eq:momN1}) is the presence of the gravitomagnetic term. In order to focus on the consequences provoked by this additional term, we remember that according to our hypothesis the gravitational field is axially symmetric, hence the gravitomagnetic field can be written in the cylindrical coordinate system introduced in the previous section in the form $\mb B= B_{r} \mb u_{r}+B_{z} \mb u_{z}$. We are interested in  circular orbits of test masses, hence we  write the velocity in the form {$\mb v=\omega r \mb u_{\varphi}$}.  The  component of Eq. (\ref{eq:lor221}) parallel to symmetry and rotation axis $z$ reads
\beq
a_{z}-\frac 2 c v_{\varphi} B_{r}=g_{z}. \label{eq:az}
\eeq
We see that, thanks to the presence of the gravitomagnetic term, non equatorial orbits ($a_{z}=0$) are possible:  this is a first remarkable difference from the Newtonian case, as pointed out by \citet{bonnor1977rotating}. In addition, Eq. (\ref{eq:az}) suggests that the effects of the gravitomagnetic field are \textit{comparable} to the Newtonian field, contrary to the common belief, deriving from the study of other physical systems, where post-Newtonian effects are always smaller than Newtonian ones: for instance, this is the case of the Solar System. \textcolor{black}{However, this is not always true: for instance, if we consider a uniformly rotating hollow homogeneous sphere, the gravitomagnetic field is constant within the sphere (see e.g.  \citet{ciufolini2003gravitomagnetic}),  while  \textit{the corresponding gravitational field is null:} this shows that it is not generally true that gravitomagnetic fields are always smaller than the Newtonian ones.}
If we assume the acceleration of stars along the rotation axis are negligible,  Eq. (\ref{eq:az}) \textit{can be used to measure gravitomagnetic effects.} {Notice that this statement is a natural consequence of the model the we considered, i.e. an axisymmetric fluid of dust: this conclusion is not necessarily true if more complex models are considered, including for instance pressure or baryonic effects.}

Following the discussion above on generalized vorticity, we might expect a natural magnitude of the  gravitomagnetic field to be that of the fluid vorticity. 

Let us evaluate the impact of the gravitomagnetic field on the rotation velocity; to this end,  we may consider
the $r$ component in the momentum equation (\ref{eq:lor221}), which turns out to be
\beq
a_{r}+\frac 2 c v_{\varphi}B_{z}=g_{r}.  \label{eq:ar}
\eeq
We suppose that test particles are moving on circles in planes orthogonal to the symmetry axis, we have $a_{z}=0, a_{r}=-\omega^{2}r$ and $v_{\varphi}=\omega r$, hence
\beq
-\omega^{2}r+\frac 2 c \omega r B_{z}=g_{r}  \label{eq:ar1}
\eeq

Now, we remember that $\mb g=-\bm \nabla \Psi$, so: $\displaystyle g_{r}=-\ppar{\Psi}{r}, g_{z}=-\ppar{\Psi}{z}$. Hence, from Eqs. (\ref{eq:az},\ref{eq:ar1}), we obtain
\begin{eqnarray}
\ppar{\Psi}{r} & = & \omega^{2}r-\frac 2 c \omega r B_{z}, \label{eq:lap1}  \\
\ppar{\Psi}{z} & = & \frac 2 c \omega r B_{r}, \label{eq:lap2}  
\end{eqnarray}
A comparison with the corresponding Newtonian equations (\ref{eq:Phiz}) and (\ref{eq:Phir}) emphasizes the role of the gravitomagnetic field. As we said before, Eq. (\ref{eq:Phir})  relates  rotation velocities and gravitational potential: we see that the non-Newtonian effect changes this equation, and it is expected that the link between mass distribution and rotation velocities is equally modified, as we are going to show. Substituting from Eqs. (\ref{eq:lap1})-(\ref{eq:lap2}) into the Poisson equation (\ref{eq:poisson1} ) and using equation (\ref{eq:Laplacian1}), we get the following expression
\begin{widetext}
\beq
2\omega^{2}+2\omega \ppar{\omega}{r}r-\frac 4 c \omega B_{z}-\frac 2 c \ppar{\omega}{r}r B_{z}-\frac 2 c \omega r \ppar{B_{z}}{r}+\frac 2 c \ppar{\omega}{z}r B_{r}+\frac 2 c \omega r \ppar{B_{r}}{z}=4\pi G \rho, \label{eq:rhoGM}
\eeq
\end{widetext}
\textcolor{black}{This equation can be directly derived in vector notation from  the momentum equation in stationary conditions (\ref{eq:mom2a}), and taking into account the Amp\`ere law for the gravitomagnetic field (see e.g. \citet{Ruggiero:2002hz,Mashhoon:2003ax}):
\beq
 \bm \nabla \wedge \mb B=\frac{8\pi G}{c}\mb j, \label{eq:ampere}
\eeq
and we obtain:
\beq
4\pi G \rho\left(1-4\frac{v^{2}}{c^{2}} \right)+\frac{2}{c}\mb B \cdot \bm \Omega=-\bm \nabla \cdot \left[\left(\mb v \cdot \bm \nabla \right)\mb v\right]. \label{eq:poissmod00}
\eeq
Since we are working at linear order in $v/c$, we can neglect the quadratic term and write
\beq
4\pi G \rho+\frac{2}{c}\mb B \cdot \bm \Omega=-\bm \nabla \cdot \left[\left(\mb v \cdot \bm \nabla \right)\mb v\right]. \label{eq:poissmod}
\eeq
For comparison, the same equation without gravitomagnetic field reads
\beq
4\pi G \rho=-\bm \nabla \cdot \left[\left(\mb v \cdot \bm \nabla \right)\mb v\right], \label{eq:poissmod0}
\eeq
which corresponds to Eq. (\ref{eq:rhoomega}).\\
The meaning of Eq. (\ref{eq:poissmod0}) is easy to understand if we take into account the momentum equation (\ref{eq:mom11}): in fact, it corresponds to Gauss' law for the gravitational field.\\
Then, from Eq. (\ref{eq:poissmod}) we see  that the coupling between the gravitomagnetic field and the fluid vorticity influences  the local relation between density and the fluid velocity.  We can evaluate the impact of the gravitomagnetic field on the density profile  setting $\displaystyle4\pi G \rho_{N}=-\bm \nabla \cdot \left[\left(\mb v \cdot \bm \nabla \right)\mb v\right]$, where $\rho_{N}$ is   the (Newtonian) density obtained without taking into account the  general relativistic effects. Then, we may set $\rho=\rho_{N} + \delta \rho$, where $\delta \rho$ is the extra density due to the coupling between the gravitomagnetic field and the fluid vorticity. We get
\beq
 \delta \rho=-\frac{1}{2\pi G c}\mb B \cdot \bm \Omega. \label{eq:posissondelta}
\eeq
For instance, in the region where the rotation curves are flat, we may set $v=\omega r \simeq \mathrm{constant}$. Accordingly, in the equatorial plane ($B_{r}=0$, $\Omega_{z}=\omega$), we obtain
\beq
\delta \rho =-\frac{1}{2\pi G c} B_{z} \omega \label{eq:posissondelta1}
\eeq
Remember that, according to Eq. (\ref{eq:rhoomega}), in this regime the Newtonian density is null: accordingly, we see that 
a gravitomagnetic field antiparallel to the $z$ contributes to define the matter density.\\
Similar conclusions about the impact of the additional degree of freedom due to the gravitomagnetic field were obtained in previous works \cite{Cooperstock:2006dt,Balasin:2006cg,crosta2020testing,Astesiano:2021ren}  studying suitable solutions of Einstein equations.\\
Our  simple argument, which rests upon few reasonable hypotheses and does not require a model for the description of the galaxy,  suggests that the gravitomagnetic field can be relevant to better estimate the impact of dark matter on the galactic rotation curves: in fact, Eq. (\ref{eq:posissondelta}) shows that part of the missing mass density needed to fit the rotation curves could  derive from general relativistic effects.}

\section{Conclusions} \label{sec:conc}

 In this paper, we have taken a quite general approach to show  the relevance of the gravitomagnetic effects in galactic dynamics. In our simplified model we consider the stars to behave as dust particles in a fluid; we expect that, far from the galactic center, the dynamics of the fluid is determined by the background spacetime that is an axially symmetric and stationary solution of Einstein's equations in the weak-field and slow-motion limit. We do not need to know the explicit form of this spacetime metric because, in any case, the gravitomagnetic formalism can be applied.

Gravitomagnetic fields could be relevant to the evolution of rotating structures like galaxies: in fact, dissipative processes in the early phase of galaxy formation are expected and these can generate fluid vorticity in protogalactic clouds (see e.g. \citet{silk1993dissipative,wang1984role}). In later evolution of galaxies, the conservation of the sum of fluid vorticity and the gravitomagnetic Larmor frequency can give rise to non-Keplerian rotation velocity curves.  {In this regard, it is useful to remember that the role and the generation of angular momentum, which is source of gravitomagnetic effects, is poorly understood: recently,  it has been showed that angular momentum can be generated on unexpectedly large scales \cite{wang2021possible}.}

{In addition, we stressed that the analogy between electromagnetism and linear gravity is limited: in non-stationary conditions,  non-Maxwellian terms appear that cannot be neglected. As a consequence in this regime the fluid vorticity is related not only to gravitomagnetic effects but, more in general, to post-Newtonian effects.}

By rephrasing an old argument by Bonnor, we showed that diverse circular orbits in planes orthogonal to the rotation axis are allowed thanks to the presence of the gravitomagnetic force that balances the Newtonian force in the direction of the rotation axis. But in Newtonian gravity the motion is the same in every plane orthogonal to the symmetry axis and there can be no variation of density in the axial direction. This raises the possibility that rotational velocities of galaxies can have a dependency on axial distance from the equatorial plane, which is contrary to what is currently assumed. {It is important to emphasise that if the galaxy can be modelled as an axisymmetric fluid of dust, Eq. (\ref{eq:az}) suggests that non equatorial circular orbits require gravitomagnetic effects of the same order as Newtonian ones: we suggest that this fact can be considered as new test of General Relativity.} \textcolor{black}{We showed that velocity, gravitomagnetic field and  mass density are connected to  the fluid vorticity:
namely, there is a coupling between the gravitomagnetic field and the fluid vorticity which influences  the local relation between density and the fluid velocity.}  {In  particular, in the Newtonian case, where this coupling is not present (see e.g. Eq. (\ref{eq:rhoomega})), the local definition of mass density seems to be meaningless in the region  where the flatness of the rotation curves is observed: this fact suggests that, in this region, the contribution of the gravitomagnetic field becomes important.}\\

 \indent In summary, our heuristic approach suggests that the gravitomagnetic field may play a relevant role in understanding galactic dynamics \textcolor{black}{as it allows for an extra rotational degree of freedom associated with  the fluid vorticity}. \textcolor{black}{It is important to emphasise that  we do not claim that this approach can eliminate the need for Dark Matter.}  {Since Dark Matter plays a central role in modern cosmology (and the rotation curves of galaxies are only one of many motivations for its existence) we believe that the physics presented in  this paper will help us to better understand the actual impact of Dark Matter: \textcolor{black}{our simple argument suggests, in fact, that an extra contribution to the matter density can derive from the coupling between the gravitomagnetic field and the fluid vorticity.}
 A more sophisticated analysis {based on a numerical approach (some numerical codes that can be used to simulate gravitomagnetism are discussed by \citet{Adamek:2020jmr})} is required to understand the details of the role that gravitomagnetism  has in the formation of galaxies, however we hope that our discussion will motivate further attention to this topic.

 {\textbf{Note Added.} After the submission of this paper to Classical and Quantum Gravity, two papers were published which focused on the role of general relativity and, in particular, of the gravitomagnetic effects on the rotation curves. \citet{Ciotti:2022inn} concludes, after a detailed analysis based on  analytical models, that GR effects  cannot compensate by any detectable amount the Keplerian fall of the rotational velocity. We note also that a simple application of Gauss' law and Ampere's circuital law to an annular element of a disk galaxy shows that the ratio of the axial components of the Newtonian, $g$, to Lorentz (i.e. gravitomagnetic) $L$, accelerations is $L/g \simeq v^2/c^2$ which supports Ciotti's analysis. However \citet{Astesiano:2022ozl} obtain a different result, starting from exact solutions of Einstein equations for stationary axisymmetric systems, thus showing that GR may have a relevant impact in understanding galactic dynamics,  since it  introduces an additional degree of freedom with respect to the Newtonian case. In particular, in the latter paper it is shown that  the strong gravitomagnetic limit can provide contributions of the same order of the Newtonian ones. Accordingly, the debate is still open and our paper can contribute to add new elements to the discussion.}

\bibliographystyle{apsrev4-1}
\bibliography{GEM_fluid}

\end{document}